# Modeling rough surfaces with Lorentz equations.

Akande R.O*, Oyewande O.E

*Theoretical Physics Group,*

*Department of Physics, University of Ibadan, Ibadan, Nigeria.*

telleverybodythat@gmail.com

**Abstract**

Surfaces sputtered by ion beam bombardment have been known to exhibit patterns whose behavior is modeled with stochastic partial differential equations. However, we apply a new approach by the use of the famous Lorentz equations to simulate and predict such patterns. It has been earlier reported that at early times, during sputtering, surface displays a chaotic pattern, with stable domains that nucleate and grow linearly in time until ripples domains of two different orientations are formed. The numerical solutions of the Lorentz model, being an unstable and chaotic model, give a pattern similar to modern surface evolution simulations. The ultimate goal was to predict the most common surface morphology by constantly varying the parameters of the Lorentz model and study the effects on the simulated surface patterns. Almost all of the recent experimentally observed and theoretically predicted formations are accurately obtainable with this concept.

# 1. Introduction

Surface sputtering is a process by which materials are removed from the surface of a solid through the impact of energetic particles. Sputtering can be used for surface analysis, depth profiling, surface cleaning, micromachining, deposition, surface coating, semiconductor doping, etching, magnetic storage technology, design of nanostructures on a surface and many more [1][2]. The sputtering process creates patterns on the surface at nanometer length scales, hence there is the need to study the surface morphology for knowledge and control of the pattern of these nanostructures exhibited on the surface when it is sputtered [2].

On the other hand, the famous Lorentz equations are one of the classical models of nonlinear dynamics and chaos [3]. They have been known to generate chaotic solutions and displays when plotted. It has been reported by [4] that at early times, during sputtering, surface displays a chaotic pattern, with stable domains that nucleate and grow linearly in time until ripples domains of two different orientations are formed [4]. Therefore, in this paper, we try to simulate the chaotic surface patterns by introducing the chaotic behavior of Lorentz equations.

# 2. Methodology

During sputtering, the surface is randomly sputtered and so the atoms become projected and re-deposited elsewhere. This is the surface morphological process where the surface evolves constantly. The process eventually subsides after sputtering and the overall surface is made up of dots, cones and the many other shapes resulting from the coalescence of these primary shapes. The morphological process that leads to these patterns is chaotic in nature and has been likened, in this paper, to the chaotic behavior of Lorentz equations (model) given in equation (1) below.

$$\left.\begin{aligned} \dot{x} &= \sigma(y - x) \\ \dot{y} &= rx - y - xz \\ \dot{z} &= xy - bz \end{aligned}\right\} \ldots\ldots(1)$$

$$\sigma = 10, r = 28, b = \frac{8}{3}$$

The $\sigma$ is the Prandtl number, b is a geometric parameter and r is the Rayleigh's number in the unit of the critical Rayleigh number. After solving each of equation (1), the difference of subsequent values, equation (2), of each solution of (1) were then arranged in a matrix form.

$$\left.\begin{array}{l}\delta_x = x_{i+1} - x_i \\ \delta_y = y_{i+1} - y_i \\ \delta_z = z_{i+1} - z_i\end{array}\right\}\ldots\ldots\ldots(2)$$

$y(0) = x(0) = z(0) = 0.$

The solution, of equation (1), was of course, numerically done with 4$^{th}$ order Runge-Kutta method with Δt=0.005. Initially, the number of iterations was 10000, and a group of 100x100 matrixes was generated for each of equation (1) starting from the first value to the last. The values of $\sigma$, r and b were varied for different iterations. The changes in the resulting surface plots were significant and impressive. It poses the possibility of predicting the surface patterns observed in most experiments.

## 3. Results

The surface plot of each group gave figures (1)-(3) below. Relating the observed peaks (derived from a fairly different origin) to surface morphology, has given way to the establishment of the similarity that exists between the behaviors of Lorentz equations and surface patterns. The peaks in figures (1)-(3) can be viewed as the cones observed in sputtering experiments which are also the grown lattice points. The holes, below the origin of the axes, are the sputtered points. The dynamics of the chaotic behavior of the Lorentz equations has obviously led to the generated surface patterns. The Lorentz chaotic pattern is being reflected in the consequent pattern on the surface. In fact, the surface pattern has even been predicted by the degree of chaos the Lorentz equations exhibit due to its initial values and the values of $\sigma$, r and b. Figures (1) – (3) are generated with the values $\sigma = 10$, $r = 28$, $b = \dfrac{8}{3}$

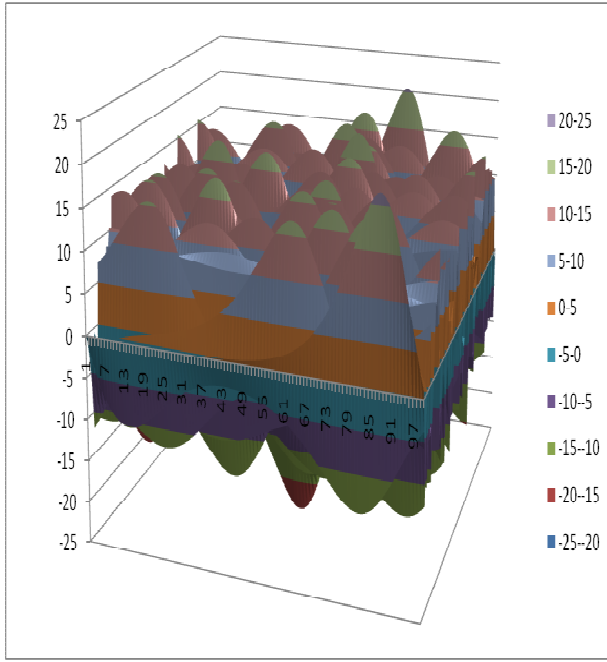

Fig(1): 3-D surface plot of solutions of $\delta_x$

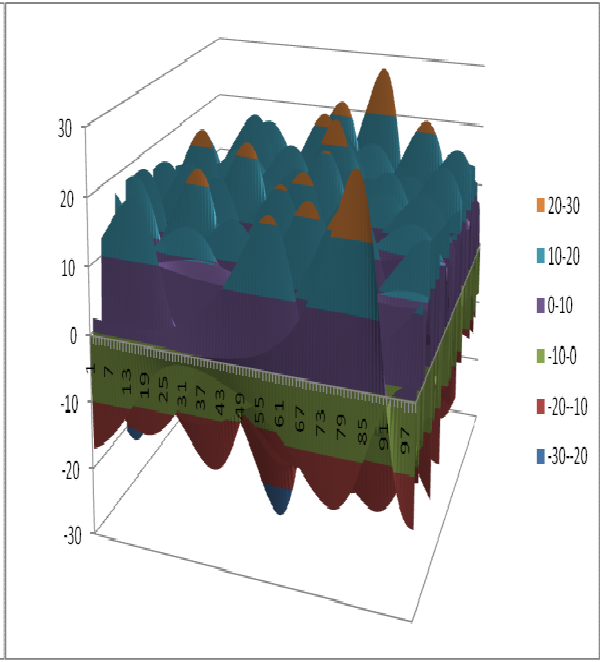

Fig(2): 3-D surface plot of solutions of $\delta_y$.

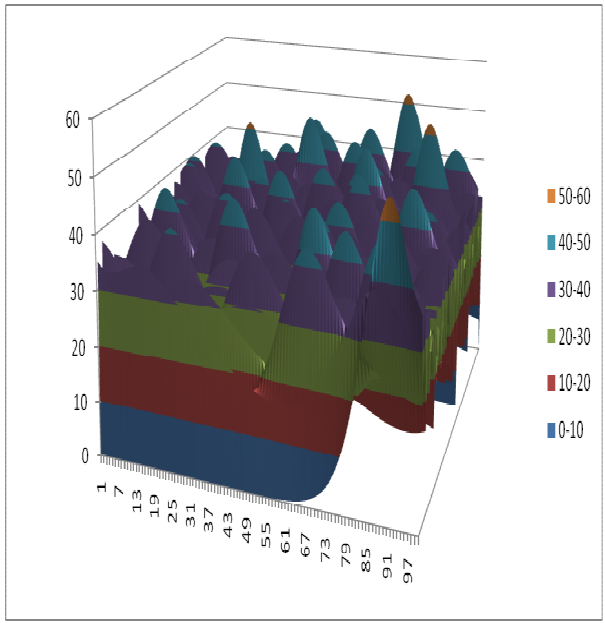

Fig(3): 3-D surface plot of solutions of $\delta_z$.

Figures (4)-(6) are solutions of equation (1) generated with $\sigma = 16, r = 45.92, b = 4$

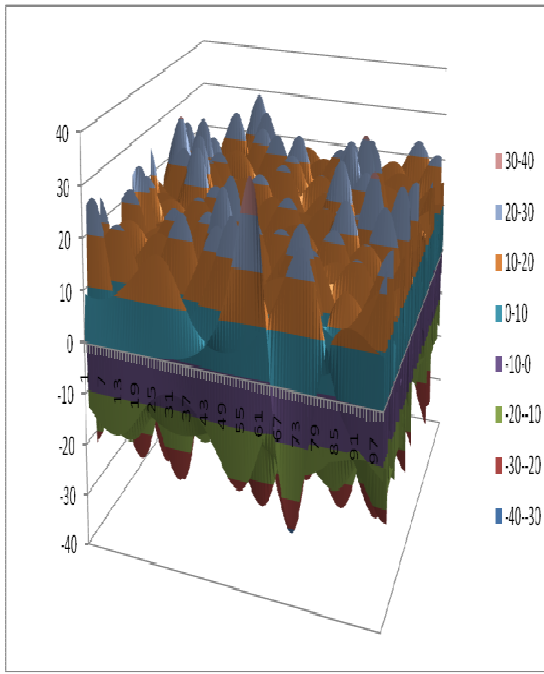
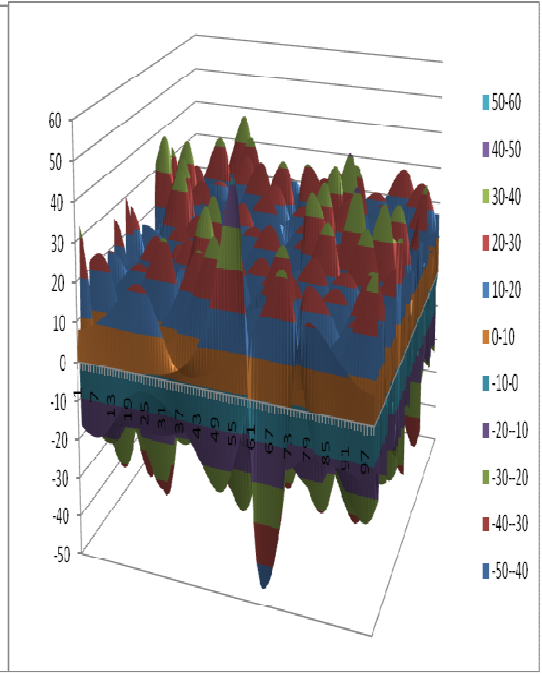

Fig(4): 3-D surface plot of solutions of $\delta_x$.    Fig(5): 3-D surface plot of solutions of $\delta_y$.

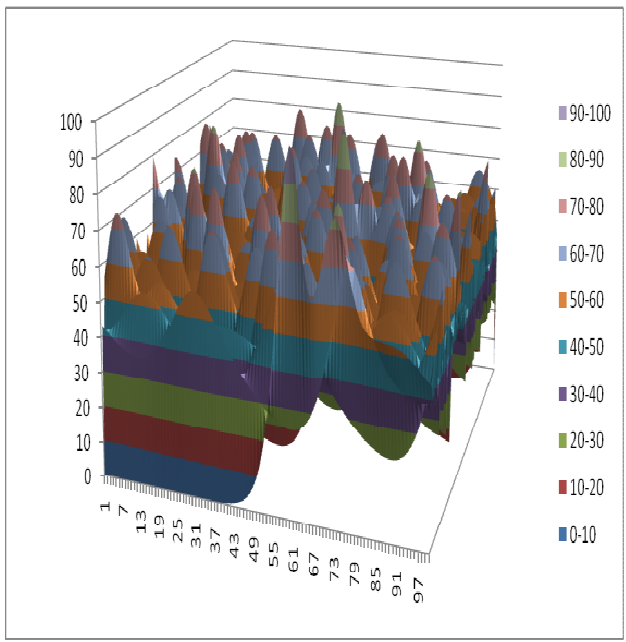

Fig(6): 3-D surface plot of solutions of $\delta_z$.

Now there is the need to study the behavior of this pattern with the period t of the Lorentz model, given the model posses an unstable (rather chaotic) periodic orbit and attractor [5]. Shown below in figures (7)-(10) are the plots of t against each of $\delta_x$, $\delta_y$ and $\delta_z$. In each plot, there exists a cone-shaped pattern, a form of bifurcation, that can traced to the sudden and simultaneous rise and fall in the solutions of equation (2). In the surface morphological pattern context, it is meaningful to say that these points are the cone (and near cone) shapes in the plots of figures (5-7). That is the points corresponding to the heaps observed in the surface plots of figures (1-3). As it is shown in figure (10), the heaps (or cones) of t against $\delta_x$, $\delta_y$ and $\delta_z$ plots merge at several points. This indicates the similarity in the behavior experienced on both axes.

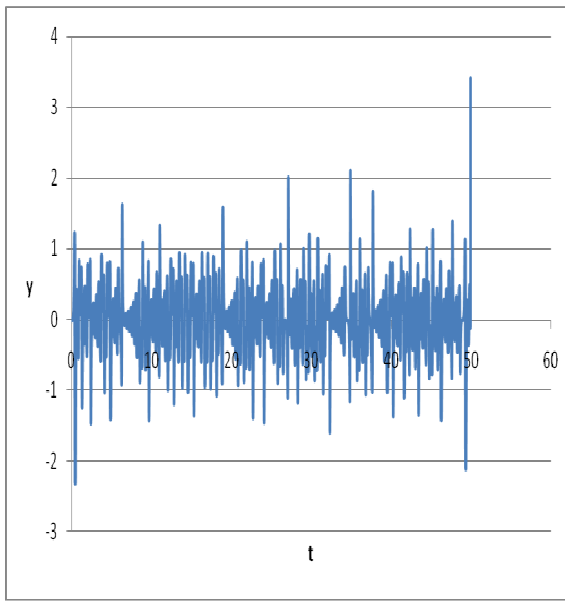
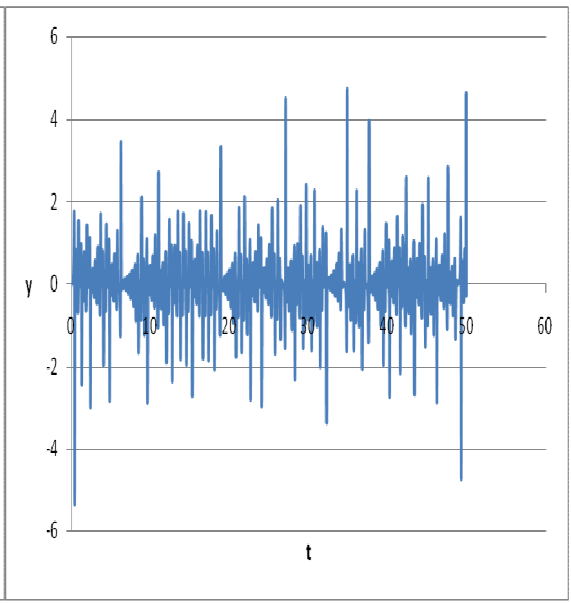

Fig(7): t vs $\delta_x$.                    Fig(8): t vs $\delta_y$

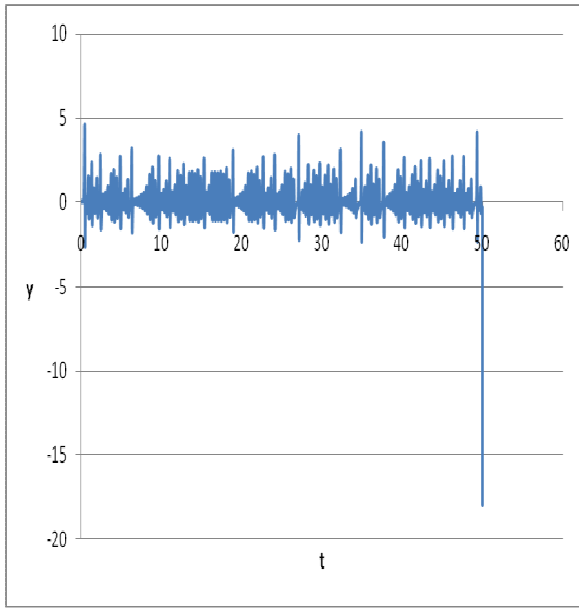
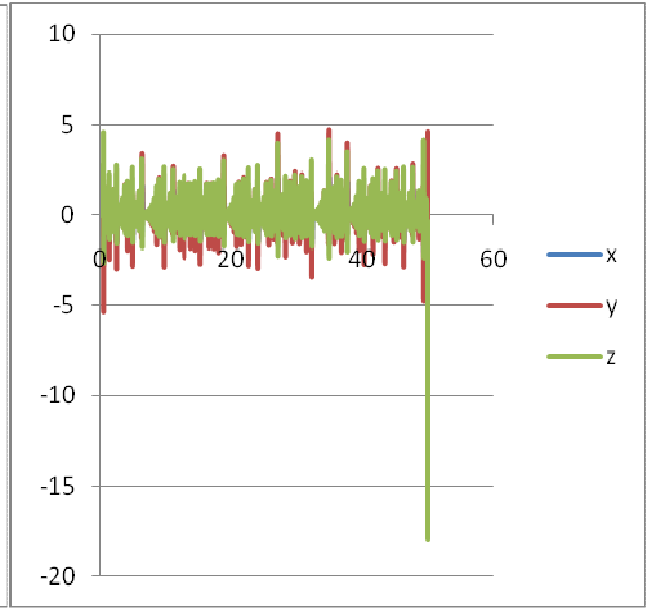

Fig(9): t vs $\delta_z$.  Fig(10): Plot of t vs $\delta_x$, $\delta_y$, $\delta_z$.

## 4. Conclusions

From the graphs in the previous section, it can be observed that the morphological patterns on most surface morphological processes are indeed not always unique but always similar in effect. This is confirmation of earlier work on morphological processes that the prediction of surface morphological processes does not always require detail atom-to-atom interaction, hence, the modeling of the surface pattern with some stochastic partial differential equations such as Bradley-Harper, Kardar-Parisi-Zhang , Cuerno-Barabasi and many other models [4]. This similarity in surface behavior is the origin of the word 'pattern'. These patterns are exhibited, when modeled with the Lorentz model, in a range of similar initial values therefore, limiting some patterns into a range of initial values.

In this paper, it is assumed that the formation of ripples and cones are due to disordered growth of the deposition sites. This can also be effectively predicted via the application of the solutions of equation (2), which is a solution of the Lorenz model. The degree of chaos within the solution of the model, due to the initial values and that of the parameters, has a significant effect in the resultant exhibited surface pattern.

# Bibliography

[1] Barabasi, A.-L., & Stanley, H. E. (1995). *Fractal Concepts in Surface Growth.* Cambridge: Cambridge University Press.

[2] Bradley, R. M., & Harper, J. M. (1988). *J. Vac. Sci. Technol. A , 6*, 2390.

[3] Giacomini, H. and Neukirch, S. C.N.R.S Laboratoire de Mathematiques et de Physique Theorique. France.

[4] Makeev, M., Cuerno, R. and Barabasi, A. (2008). arxiv: [cond-mat.matrl-sci], 0007354v1.

[5] Landau R.H, & Manuel Jose. Computational Physics: Wiley-VCH Verlag GmbH & Co. KGaA.